\documentclass[%
reprint,
superscriptaddress,
footinbib,
 amsmath,amssymb,
aps,
floatfix,
]{revtex4-2}

\usepackage{graphicx}
\usepackage{dcolumn}
\usepackage{bm,soul}
\usepackage{units}

\usepackage[colorlinks,allcolors=blue,hyperindex,breaklinks]{hyperref}
\usepackage{qcircuit}

\newcommand{\ket}[1]{\ensuremath{\vert{#1\rangle}}} 
\newcommand{\bra}[1]{\ensuremath{{\langle #1}\vert}}
\newcommand{\braket}[2]{\ensuremath{{\langle #1}\vert{#2 \rangle}}}
\newcommand{\ketbra}[2]{\ensuremath{|{#1 \rangle}{\langle #2}|}}
\newcommand{\op}[1]{\hat{#1}}

\newcommand{\markrev}[1]{\textcolor{black}{#1}}

\newcommand{\I}{\text{i}}
\newcommand{\E}{\text{e}}

\begin{document}

\title{Protective measurements of photon polarization using a temporal pointer}

\author{Meng-Wei Chen}
\author{Owen Young}
\affiliation{Department of Physics, Reed College, 3203 SE Woodstock Blvd., Portland, Oregon 97202, USA}

\author{Maximilian Schlosshauer}
\affiliation{Department of Physics, University of Portland, 5000 N Willamette Blvd., Portland, Oregon 97203, USA}

\author{M. Beck}
 \email{beckm@reed.edu}
\affiliation{Department of Physics, Reed College, 3203 SE Woodstock Blvd., Portland, Oregon 97202, USA}

\date{\today}

\begin{abstract}
We experimentally demonstrate protective measurements by weakly coupling the polarization of a single-photon-level field to a measurement pointer that corresponds to the arrival time of the photon. By using an optical loop, we implement a variable, controlled number (1--9) of protection and measurement stages. We demonstrate the measurement of expectation values of photon polarization by measuring arrival times while simultaneously protecting the polarization state. No knowledge of the initial photon state is required or available in our experiment, demonstrating that protective measurements provide a genuine information gain that cannot simply be reduced to \emph{a priori} information about the protection procedure.\\[.2cm]
Journal reference: \emph{Phys.\ Rev.\ A }{\bf 108}, 022420 (2023), doi: \href{https://doi.org/10.1103/PhysRevA.108.022420}{\texttt{10.1103/PhysRevA.108.022420}}

\end{abstract}

\maketitle

\section{\label{sec:intro}Introduction}

Weak quantum measurements with postselection have become an essential tool in quantum metrology and the study of quantum phenomena \cite{Aharonov:1988:mz,Tamir:2013:za,Dressel:2014:uu,ArvidssonShukur:2020:az}. Protective measurements (PM) 
\cite{Aharonov:1993:qa,Aharonov:1993:jm,Anandan:1993:uu,Dass:1999:az,Vaidman:2009:po,Gao:2014:cu,Genovese:2017:zz,Piacentini:2017:oo,rebufello_2021}
are a type of weak measurement in which a procedure is added that prevents the state of the system from changing appreciably during the measurement. The result of the PM on a single system is the expectation value of an arbitrary observable. 
PMs can provide better estimates of expectation values (lower uncertainties) than could be achieved, using comparable resources, with strong measurements (SMs) on an ensemble \cite{Piacentini:2017:oo}. Measurements with state protection also provide an avenue for beating the Cram{\'e}r-Rao bound \cite{Zhang:2020:aa}.

A common version of a PM uses a state-protection procedure based on the quantum Zeno effect \cite{Misra:1977:aa,Itano:1990:lm,Home:1997:za,Virzi:2022:aa} and is referred to as a Zeno PM \cite{Aharonov:1993:jm}. The system interacts repeatedly and weakly with an apparatus, and between each interaction the system is projected back onto its initial state, thus protecting it. We refer to this combination of a weak interaction followed by a protection step as a Zeno stage. The Zeno stages  amount to a series of identical weak measurements on the same system, with the pre- and postselected states being the same, such that the weak value \cite{Aharonov:1988:mz,Dressel:2014:uu} reduces to the expectation value. The performance improves as the number of Zeno stages is increased while the interaction strength for each stage is weakened \cite{Aharonov:1993:jm,Piacentini:2017:oo}. The expectation value is read off from the accumulated pointer shift after all Zeno stages. This scheme does not require knowledge of the initial state. It is only needed that the state preparation and protection procedures project onto the same (but potentially unknown) state. 
The first, and previously only, experimental realization of a PM, based on the Zeno scheme, was reported by Piacentini \emph{et al.\ }\cite{Piacentini:2017:oo,rebufello_2021}, using seven Zeno stages implemented as individual units.

Here we report on the experimental realization of a Zeno PM of photon polarization in which the photons are cycled through a loop constructed from an optical fiber, such that a given photon will pass an adjustable number of times through the same Zeno stage \cite{Schlosshauer:2018:xx}. 
This loop configuration is advantageous because (i) we do not need to physically implement several individual Zeno stages, and (ii) we can easily control and adjust the number of stages. Furthermore, the particular state-protection method we use does not require, and indeed precludes, knowledge of the initial photon state, even in principle. Thus, our experiment explicitly demonstrates that a Zeno PM can provide genuine information gain. (If the protection necessitated knowledge of the state, the measured expectation value could, instead, be simply calculated from this information.)

In contrast to the spatial pointer shift used in Ref.~\cite{Piacentini:2017:oo}, we encode the measurement result in a temporal degree of freedom. The temporal pointer shift is generated by a polarization-dependent differential group delay (DGD) between the horizontal and vertical polarization directions, and the shift of the pointer is obtained by measuring photon arrival times. An advantage of a temporal pointer is that it is compatible with the use of single-mode fibers, which eliminate the spatial degrees of freedom available to free-space optics. The use of DGD to realize general weak measurements of photon polarization was described in Ref.~\cite{Brunner:2003:az}, and realizations of such weak (but not protective) measurements in the time domain were reported in Refs.~\cite{Brunner:2004:aa,Wang:2006:un}. 

This paper is organized as follows. Section~\ref{sec:theory} provides the theory of a Zeno PM, applied to our experimental setting. Section~\ref{sec:expt} describes  our experimental apparatus and data acquisition. Experimental results are presented in Sec.~\ref{sec:results}. We offer a concluding discussion in Sec.~\ref{sec:conclusions}.

\section{\label{sec:theory}Theory}

A measurement interaction of duration $\Delta t$ between a quantum system $\mathcal{S}$ and an apparatus $\mathcal{A}$ may be described in terms of a unitary operation 
\begin{equation}\label{eq:H}
\op{U} = \exp\left[-\I \xi (\op{O} \otimes \op{P})\Delta t\right],
\end{equation}
where $\xi$ is the coupling strength, $\op{O}$ is an arbitrary observable of $\mathcal{S}$, and $\op{P}$ generates the shift of the pointer of $\mathcal{A}$ in the conjugate variable $q$. We make the usual (and experimentally appropriate) assumption that $\mathcal{A}$ starts out in a Gaussian of width $\Delta$ in $q$, $\Psi(q) = \frac{1}{\sqrt{\Delta} (2\pi)^{1/4}} \exp\left( -\frac{q^2}{4\Delta^2}\right)$. In a weak measurement, the coupling strength is small in the sense that the shifts of the pointer  induced by Eq.~\eqref{eq:H} for the different eigenvalues of $\op{O}$ are small compared to the width $\Delta$. A weak measurement consists of three stages: (i) the preparation (preselection) of the initial state $\ket{\psi_0}$ of $\mathcal{S}$, (ii) the interaction described by Eq.~\eqref{eq:H}, and (iii) the postselection of $\mathcal{S}$ in a state $\ket{\psi_p}$. Then  one can show \cite{Aharonov:1988:mz} that following postselection, the pointer of $\mathcal{A}$ shifts by an amount $\xi O_w \Delta t$, where
\begin{equation}\label{eq:haavsg7117788aaiuxxh}
O_w = \frac{ \bra{\psi_p}\op{O}\ket{\psi_0}}{\braket{\psi_p}{\psi_0}},
\end{equation}
is called the weak value of $\op{O}$.

\begin{figure*}
\begin{equation*}
\centering
\Qcircuit @C=1em @R=.7em {
\lstick{\mathcal{S}: \ket{\psi_0}}     & \multigate{1}{U} &  \measure{\ket{\psi_0}} & \qw& \multigate{1}{U} &  \measure{\ket{\psi_0}} & \qw & \push{\cdots\quad}  & \multigate{1}{U} &  \measure{\ket{\psi_0}} & \qw \\
\lstick{\mathcal{A}}     & \ghost{U}&  \qw &\qw& \ghost{U}&  \qw & \qw & \push{\cdots\quad} & \ghost{U}&  \qw & \meterB{\langle\op{O}\rangle}
\gategroup{1}{2}{2}{3}{.7em}{--} 
\gategroup{1}{5}{2}{6}{.7em}{--} 
\gategroup{1}{9}{2}{10}{.7em}{--} 
}
\end{equation*} 
\caption{\label{fig:lm}Principle of a Zeno protective measurement. After each weak interaction with the measuring device, the system is projected back onto its initial state $\ket{\psi_0}$. Each dashed unit represents a Zeno stage.}
\end{figure*}
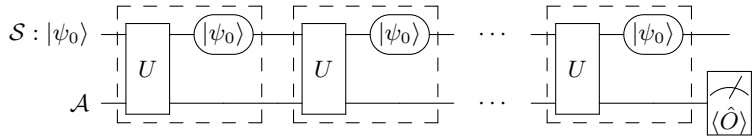

The Zeno stages in a Zeno protective measurement \cite{Aharonov:1993:jm} effectively amount to a series of identical weak measurements on the same system, with the pre- and postselected states being the same (Fig.~\ref{fig:lm}). Then the weak value~\eqref{eq:haavsg7117788aaiuxxh} reduces to the expectation value $\langle\op{O}\rangle=\bra{\psi_0}\op{O}\ket{\psi_0}$, which can be read off from the pointer shift accumulated over multiple Zeno stages. 

We now consider the case relevant to our experiment, in which the pointer shift is produced by temporal DGD. Consider the passage of a photon prepared in a polarization state $\ket{\psi_0}  = \cos\theta\ket{H}+\sin\theta e^{i\phi}\ket{V}$ through a birefringent material of length $L$. Let the group velocities be $v_{g,H}$ and $v_{g,V}$ for the ordinary ($H$ polarization) and extraordinary (polarization $V$) rays, with corresponding travel times $t_i =\frac{L}{v_{g,i}}$, $i=H,V$. 
Let the pulse entering the material be described by a Gaussian of duration $\tau_{G}$,
\begin{equation}\label{eq:1}
\mathcal{E}(t,z=0) = \mathcal{E}_0 \E^{-(t/\tau_{G})^2},
\end{equation}
and let us associate a quantum state $\ket{\phi(0)}$ with this pulse. After passage through the birefringent material, the pulses corresponding to the ordinary and extraordinary rays are
\begin{equation}
\mathcal{E}_i(t,L) = \mathcal{E}(t-t_i,z=0) = \mathcal{E}_0 \exp\left\{ -\left[\frac{t-t_i}{\tau_{G}}\right]^2\right\}, 
\end{equation}
where $i=H,V$. Define the average travel time $t_\text{avg} = [t_H+t_V]/2$ and the relative delay $\tau = t_H-t_V$. Also, express all times in terms of a dimensionless time variable $\tilde{t}=t/\tau_G$. Then we can write
\begin{align}
\mathcal{E}_H(\tilde{t},L) &= \mathcal{E}_0 \exp\left\{ -\left[\tilde{t} -(\tilde{t}_\text{avg}-\tilde{\tau}/2)\right]^2\right\}, \\
\mathcal{E}_V(\tilde{t},L) &= \mathcal{E}_0 \exp\left\{ -\left[\tilde{t} -(\tilde{t}_\text{avg}+\tilde{\tau}/2)\right]^2\right\},
\end{align}
and we denote the associated quantum states by $\ket{\phi(\pm \tau/2)}$. The interaction is assumed to be weak in the sense that $\tau \ll \tau_{G}$, i.e., $\tilde{\tau} =\tau/\tau_G \ll 1$ (this means that the separation of the temporal wavepackets associated with the orthogonal polarizations is incomplete). In analogy with Eq.~\eqref{eq:H}, we can model the relative temporal displacement $\pm \tilde{\tau}/2$ in terms of a unitary operator 
\begin{equation}\label{eq:3aaa}
\op{U}(\tilde{\tau}) = \exp\left[ -\I \frac{\tilde{\tau}}{2} \op{O} \otimes \op{A} \right],
\end{equation}
where $\op{O} = \ketbra{H}{H} - \ketbra{V}{V}$ is the linear polarization observable,  $\op{A}$ generates shifts of the temporal Gaussian wavepacket, and $\tilde{\tau} = \tau/\tau_G$ plays the role of an interaction strength. 

After passing through the birefringent material and subsequent projection onto $\ket{\psi_0}$, the (unnormalized) photon state is
\begin{equation} \label{eq:11}
\ket{\Psi_1} =  \ket{\psi_0} \bra{\psi_0}\exp\left[ -\I \frac{\tilde{\tau}}{2} \op{O} \otimes \op{A} \right] \ket{\psi_0} \ket{\phi(0)}.
\end{equation}
Since the interaction is weak, we can expand to second order in $\tilde{\tau}$ \cite{Aharonov:1993:jm}:
\begin{multline}
\bra{\psi_0}\exp\left[ - \I \frac{\tilde{\tau}}{2} \op{O} \otimes \op{A} \right] \ket{\psi_0}  \\ =
 1 - \I \frac{\tilde{\tau}}{2}\langle \op{O} \rangle \op{A} - \frac{1}{2} \left(\frac{\tilde{\tau}}{2}\right)^2 \langle \op{O} \rangle ^2 \op{A}^2 
- \frac{1}{2} \left(\frac{\tilde{\tau}}{2}\right)^2\Delta O^2 \op{A}^2,
\end{multline}
where $\langle \op{O} \rangle = \bra{\psi_0} \op{O} \ket{\psi_0}=\cos 2\theta$ and $\Delta O^2 =\langle \op{O}^2 \rangle  - \langle \op{O} \rangle ^2$ is the square of the uncertainty in $\op{O}$. Reintroducing the exponential function (and again working to second order in $\tilde{\tau}$) gives \cite{Aharonov:1993:jm}
\begin{multline}
\bra{\psi_0}\exp\left[ - \I \frac{\tilde{\tau}}{2} \op{O} \otimes \op{A} \right] \ket{\psi_0} \\ =
\exp\left[ - \I \frac{\tilde{\tau}}{2}  \langle \op{O} \rangle  \op{A} \right]\left[1- \frac{1}{2} \left(\frac{\tilde{\tau}}{2}\right)^2\Delta O^2 \op{A}^2\right].
\end{multline}
After $\ell$ Zeno stages, the final state is therefore of the form
\begin{align}\label{eq:5}
\ket{\Psi_\ell} &= \ket{\psi_0} \left[1- \frac{1}{2} \left(\frac{\tilde{\tau}}{2}\right)^2\Delta O^2 \op{A}^2\right]^\ell \notag \\ & \qquad \times \exp\left[ - \I \frac{\ell\tilde{\tau}}{2}  \langle \op{O} \rangle  \op{A} \right] \ket{\phi(0)} \notag \\ &= \ket{\psi_0} \left[1- \frac{1}{2} \left(\frac{\tilde{\tau}}{2}\right)^2\Delta O^2 \op{A}^2\right]^\ell
\ket{\phi(\ell\tilde{\tau}\langle \op{O} \rangle/2)}.
\end{align}
We see that the temporal shift  of the pulse wave packet is equal to $\ell\tilde{\tau}\langle \op{O} \rangle/2$. Thus, the birefringent material delays the arrival of the photon by $\Delta t = \ell\frac{L}{c} - \ell\left(t_\text{avg} -\frac{\tau}{2}\langle \op{O} \rangle\right)$. By measuring this delay, we can measure $\langle \op{O} \rangle$. 

The probability for the photon to survive the projections is determined by the overlap of the temporal wave packets (the bigger the overlap is, the higher the survival probability is), which is related to the ratio $\tilde{\tau} = \tau/\tau_G$ (a small $\tilde{\tau}$ implies a large overlap). By decreasing the temporal shift induced by a single pass through the birefringent material (i.e., by decreasing $\tilde{\tau}$), the probability of the photon reaching the output after $\ell$ stages can be increased. Doing so, however, will also decrease the total shift at the output, leading to greater uncertainty in the measured expectation value. To compensate, one may increase $\ell$. Specifically, by increasing $\ell$ while proportionally decreasing $\tilde{\tau}$ (i.e., keeping $\ell \tilde{\tau}$ constant to maintain a fixed amount of total pointer shift and therefore measurement resolution), the survival probability can be made arbitrarily close to unity (see also Ref.~\cite{Piacentini:2018:za}). It is important to note that while increasing $\ell$ (while keeping the measurement strength constant) will decrease the photon survival probability, the decrease grows very slowly with $\ell$, much slower than the decrease in uncertainty \cite{Piacentini:2017:oo}. 

\begin{table}
\begin{ruledtabular}
\begin{tabular}{cccccc}
$\theta$ & $\tilde{\tau}$ & $\ell$ & $\mathcal{P}$ & Pointer shift & $\langle O \rangle$ \\\hline 
$0.2\frac{\pi}{4}$ & 0.1 &10 & 0.994 & 0.951 & 0.951 \\
& 0.2 &5  & 0.988 & 0.952 & \\
& 0.5 &2  & 0.972 & 0.961 & \\\hline
$0.8\frac{\pi}{4}$ & 0.1 & 10  & 0.946 & 0.310 & 0.310 \\
& 0.2 & 5  & 0.894 & 0.313 & \\
& 0.5 & 2  & 0.753 & 0.353 & \\\hline
$1.5\frac{\pi}{4}$ & 0.1 & 10  & 0.967 & $-0.708$ & $-0.707$ \\
& 0.2 & 5  & 0.940 & $-0.712$ & \\
& 0.5 &2  & 0.859 & $-0.757$ & \\ 
\end{tabular}
\end{ruledtabular}
\caption{\label{tab:prob}Photon-survival probability $\mathcal{P}$ [Eq.~\eqref{eq:surv}] and accumulated pointer shift for different input states $\ket{\psi_0}  = \cos\theta\ket{H}+\sin\theta\ket{V}$ as a function of the measurement strength $\tilde{\tau}$ and the number $\ell$ of passes (projections). The value $\tilde{\tau}=1$ corresponds to a strong measurement that completely resolves the $H$ and $V$ components in a single pass. The last column represents the expectation value of the polarization observable $\op{O}=\ketbra{H}{H} - \ketbra{V}{V}$. }
\end{table}

We shall give a few explicit values for the photon survival probability after $\ell$ passes. This probability is given by 
\begin{equation}\label{eq:surv}
\mathcal{P} = \text{Tr} \left[ \left(\prod_{i=1}^\ell \ketbra{\psi_0}{\psi_0} \op{U}(\tilde{\tau})\right) \left[\ketbra{\psi_0}{\psi_0} \otimes \ketbra{\phi(0)}{\phi(0)} \right]\right],
\end{equation}
where $\op{U}(\tilde{\tau})$ is the evolution operator \eqref{eq:3aaa}. Table~\ref{tab:prob} shows survival probabilities that were numerically calculated from Eq.~\eqref{eq:surv} for different polarization states and measurement strengths $\tilde{\tau}$. We  chose the number $\ell$ of passes such that $\ell\tilde{\tau}=1$, so that the pointer shift falls within the range $[-1,1]$. We also compare the pointer shift with the expectation value of the polarization observable, i.e., with the value we would like to obtain from the pointer shift. The results show that already for $\ell=5$ passes we get high survival probabilities and excellent agreement between the pointer shift and expectation value. For $\ell=10$, the survival probabilities are $>94\%$. Note that these probabilities are obtained solely from the probability of a photon surviving the multiple projections. Of course, in practice additional losses will come from the fiber-optic elements in the loop. In our experiment,  these losses are $\sim\!\unit[10]{dB}$ (90\%) per loop, substantially larger than the losses due to the projections.


\section{\label{sec:expt}Experiment}
\subsection{\label{sec:app}Experimental apparatus}

\begin{figure}
\centering\includegraphics[width=8.5 cm]{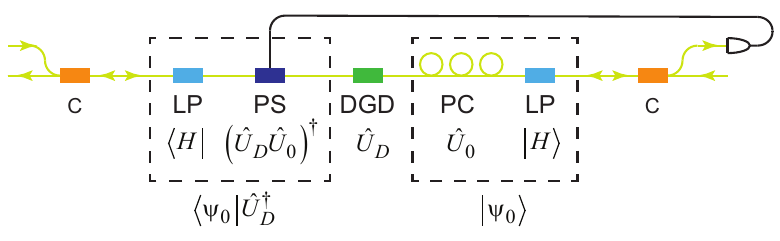}
\caption{\label{fig:stab}The experimental apparatus for a single Zeno stage of the protective measurement. Abbreviations: circulator (C), linear polarizer (LP), polarization controller (PC), differential group delay (DGD) and polarization stabilizer (PS). The circulators multiplex and demultiplex beams traveling in opposite directions. Individual photons travel right to left, while the stabilization beam travels left to right.}
\end{figure}

The experimental apparatus used for a single Zeno stage is shown in Fig.~\ref{fig:stab}. For a PM to succeed, as described above, after the system interacts weakly with the apparatus, it is necessary to project back onto the initially prepared state. To accomplish this we use a commercial polarization stabilizer (Luna POS-002). 

All of the light in our experiments is derived from a \unit[1540]{nm}-laser, and it propagates through single-mode optical fibers. We use circulators to multiplex and demultiplex beams that propagate in opposite directions. In Fig.~\ref{fig:stab} the single-photon-level signal beam propagates from right to left, while a continuous-wave (cw) laser beam travels from left to right. First, we consider the single-photon-level beam that travels from right to left. It passes through a circulator, then a linear polarizer (LP) that projects onto $\ket{H}$. A manual polarization controller (PC; three loops of fiber that can be rotated) can be configured to create any state of polarization by implementing a unitary transformation $\hat{U}_0$. The combination of the LP and the PC prepares the state $\ket{\psi_0}=\hat{U}_0\ket{H}$. The photons then pass through the DGD, which imparts \unit[0.5]{ns} of DGD between the fast ($\ket{V}$) and slow ($\ket{H}$) axes of the fiber.  However, the DGD also shifts the relative phase of the $\ket{H}$ and $\ket{V}$ states, affecting the polarization state as well. This state change is described by the unitary transformation $\hat{U}_D$. The DGD is implemented by a long ($\unit[\sim 250]{m}$) length of birefringent fiber, and the phase shift it imparts is extremely sensitive to environmental conditions. We passively stabilize the phase shift by temperature stabilizing this fiber to less than $\unit[ 0.02]{{^\circ}C}$, but, despite this, the phase drifts slowly in time, necessitating active stabilization. 

The photons pass through a polarization stabilizer (PS) before passing through a second linear polarizer that projects onto $\ket{H}$ and a second circulator. 
The PS implements the unitary transformation ${\left( \hat{U}_D \hat{U}_0 \right)}^ \dagger$. The total transformation between the linear polarizers is thus equal to the identity, so photons that are transmitted by the first polarizer will also be transmitted by the second. This means that the combination of the PS and the second linear polarizer effectively undoes the phase shift induced by the DGD, and then projects back onto the initial state $\ket{\psi_0}$\markrev{, as illustrated in Fig.~\ref{fig:stab}}. The PS implements the proper transformation via feedback derived from the cw beam that propagates between the polarizers in the opposite direction of the single-photon level field. The PS automatically adjusts the phase shift to maximize the intensity of the cw beam transmitted by the second polarizer, guaranteeing that the total polarization transformation between the polarizers is equal to the identity operation. We note that there are also losses in the components between the polarizers, but as long as these losses are not polarization dependent they affect the overall transmission and do not affect the polarization transformations. We have chosen components with an eye toward minimizing polarization-dependent losses, and are unable to observe any such losses in our experiments.

To characterize the performance of our PS, we can insert a beamsplitter between the PC and the LP in Fig.~\ref{fig:stab}, and pick off a small portion of the stabilization beam to send to a polarization analyzer. (This beam splitter is present only during characterization, not during our PM experiments.) We find that when the PS is on, the Stokes vector that describes the classical polarization has an approximately Gaussian distribution, if it is expressed as a point on the Poincar\'e sphere. When the polarization is nearly linear (located near the equator of the Poincar\'e sphere), the angular spread in this distribution is found to have a standard deviation of $\unit[0.08]{rad}$. It is reasonable to assume that the difference between the Bloch vectors that describe the prepared quantum polarization state, and the polarization state that is projected onto, is described by this same distribution. Two polarization states whose Bloch vectors differ by $\unit[0.08]{rad}$ have a fidelity of 0.998, and we believe this is a good estimate of the fidelity between our prepared and projected states.

\markrev{The transformation $\hat{U}_0$, which produces the state $\ket{\psi_0}$ we are measuring, is stable in time.} It is the transformation $\hat{U}_D$ imparted by the lengthy DGD line that is random (indeed, it varies randomly in time). The PS implements the transformation ${\left( \hat{U}_D \hat{U}_0 \right)}^ \dagger$, and this information is, in principle, available to the experimenter. However, there is no way to separate the contributions from $\hat{U}_D$ and $\hat{U}_0$, so the experimenter gains no information about $\hat{U}_0$. The polarization stabilization thus guarantees that the state $\ket{\psi_0}$ is protected but yields no information, even in principle, about what that state is.

Despite that fact that the classical stabilization beam yields no information about $\ket{\psi_0}$, information about this state \emph{is} retained in the counterpropagating, single-photon-level beam, and this is why our PM is possible. The transformation $\hat{U}_0$ produces the state $\ket{\psi_0}$ at the input of the DGD. It is the state at this location that is protected, as described above. It is this state that experiences the temporal delay that we measure in our experiment. The phase shift in the DGD does not affect the relative time delay between the $\ket{H}$ and $\ket{V}$ components of the single-photon level pulse. The PS ensures that the relative phase of these components is adjusted so that they properly interfere at the final LP, projecting onto $\ket{\psi_0}$.  

A related point is that the PS affects only the relative phase of the two polarization components, but the operator $\op{O} = \ketbra{H}{H} - \ketbra{V}{V}$ that we are performing a PM of (the same operator that was measured in the previous PM experiment \cite{Piacentini:2017:oo}) is not sensitive to this phase. It is the relative amplitudes of $\ket{H}$ and $\ket{V}$ at the input to the DGD that the measurement is sensitive to. We would need to redesign our apparatus to measure a different polarization operator, which would be experimentally feasible, in order for it to be sensitive to the relative phase.

\begin{figure}
\centering\includegraphics[width=8.5 cm]{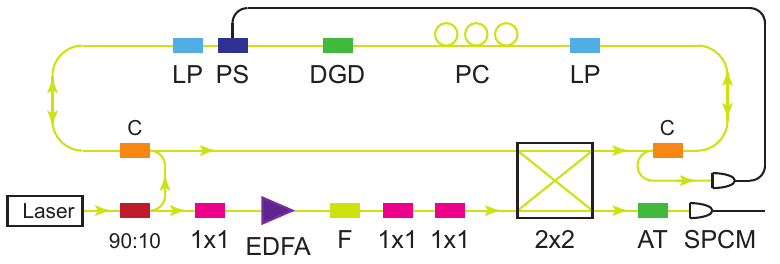}
\caption{\label{fig:app}The complete experimental apparatus. Abbreviations: erbium-doped fiber amplifier (EDFA), bandpass filter (F), circulator (C), linear polarizer (LP), polarization controller (PC), differential group delay (DGD), polarization stabilizer (PS), variable attenuator (AT) and single-photon counting module (SPCM). A $90\%$:$10\%$ splitter is denoted by 90:10. Amplitude modulators are shown as $1 \times 1$ switches, and a $2 \times 2$ switch directs light into and out of a fiber loop containing the single-Zeno-stage apparatus (Fig.~\ref{fig:stab}).}
\end{figure}

Our complete experimental apparatus is shown in Fig.~\ref{fig:app}. Ninety percent of the light from a cw, 1540-nm laser passes through an amplitude modulator ($1 \times 1$ switch), which slices out a 3-ns-long pulse. The pulse repetition rate is $\unit[50]{kHz}$. The pulse is amplified by up to $\unit[20]{dB}$ in an erbium-doped fiber amplifier, and filtered by a 100-GHz bandpass filter to eliminate amplified spontaneous emission. The pulse then passes through two more $1 \times 1$ switches in order to further decrease cw background light. A $2 \times 2$ switch directs the pulse into a fiber loop, where it propagates in the counterclockwise direction. The timing of the switch is set to allow the pulse to travel a predetermined number of times around the loop before being switched out. Inside the loop the pulse traverses the single-Zeno-stage apparatus (Fig.~\ref{fig:stab}) before once again reaching the $2 \times 2$ switch. After the pulse is switched out, it is attenuated to the single-photon level ($< 0.1$ photon per pulse) before being detected with a single-photon counting module (SPCM) with a temporal resolution of $\unit[100]{ps}$; the SPCM is gated on for a 15-ns window surrounding the arrival of the pulse. Photon arrival times are measured with a time-to-digital converter with a temporal resolution of $\unit[20]{ps}$. 

Because the pulses travel around a loop, the LP-PC combination that produces the initial state $\ket{\psi_0}$ is the same for each iteration of the PM. Thus, to high precision, this state is the same for each iteration.
 
 \subsection{\label{sec:data}Data acquisition and analysis}
 
For each data run, we fix the polarization of photons incident on the DGD using the PC and set the timing of the $2 \times 2$ switch to realize the desired number $\ell$ of loops (see Fig.~\ref{fig:app}). For each polarization setting and each $\ell$, we acquire data as follows.  For one to six loops we acquire $\sim\!750\,000$ total (signal and background) counts, and this takes about \unit[150]{s}. Above six loops the count rate drops, and the signal-to-background ratio becomes smaller, so we acquire more data in order to have sufficient signal counts. For 9 loops we acquire $\sim\!3\,500\,000$ total counts, which takes about \unit[9000]{s}.
The time-tagged photon counts are sorted into histograms, with bins denoted by measured arrival time. To acquire the background, we change the timing to move the pulse out of the recorded time window. Background counts are acquired for between 150 and \unit[900]{s}, depending on the count rate, and histogrammed using the same bins as for the signal. For each data set we calculate the total number of counts in a 2--4-ns window away from the peak in the set, where we have only background counts. We then calculate the total number of counts in the same window for the corresponding background data set. We scale the background data so that they have the same number of counts in that window as the data, and then subtract the background. We have found that this background subtraction procedure works better for subtracting some small structure in the background (likely due to the behavior of the $2 \times 2$ switch) than simply subtracting a constant, average background.

To further reduce the effect of background when calculating the statistics of the arrival times, we truncate the histograms to remove data in arrival-time bins that are away from the peak of the histogram and contain only background. 
Starting from the histogram peak, we move to shorter (longer) times, and
locate the first time at which the background-subtracted histogram has a negative value. We truncate the histogram at this point, removing all data corresponding to this and shorter (longer) arrival times. The histograms now contain only positive values, and we normalize them to obtain estimates of the probability $P_i$ for a photon to arrive in time bin $t_i$. Our truncation procedure eliminates nonphysical negative probabilities, in a consistent manner, while ensuring that the relevant probabilities are preserved. 

To set the time scale for a given $\ell$, the mean of the  distribution for photons polarized along the fast axis of the DGD 
is used to define the zero of the scale, and arrival times for other polarization settings $\mathcal{P}$ are 
expressed in terms of the relative delay $\tau(\mathcal{P}) = \bar{t}_\text{ar}(\mathcal{P}) - \bar{t}_\text{ar}(V)$, 
where $\bar{t}_\text{ar}(V)$ is the mean arrival time for photons polarized along the fast axis. 

\section{\label{sec:results}Results}

\begin{figure}
\centering\includegraphics[width=7 cm]{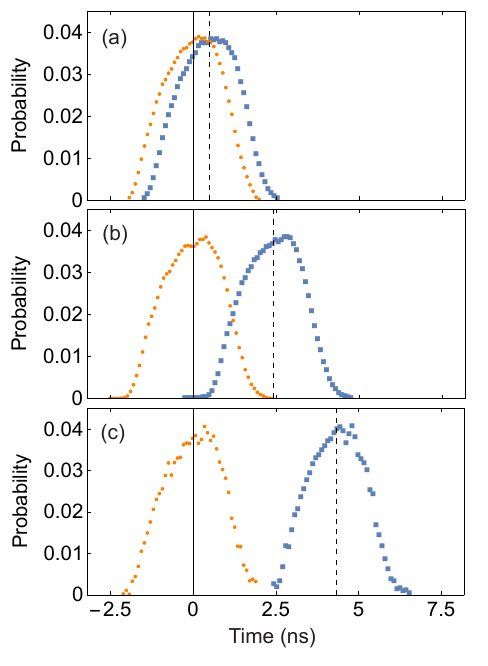}
\caption{\label{fig:1_5_9} Probabilities of photon arrival times after (a) one loop, (b) five loops, and (c) nine loops, for polarization along the fast (orange circles) and slow (blue squares) axes.
Vertical lines represent mean arrival times. 
The fast-axis polarization is defined to have a mean arrival time of 0, while the mean arrival time of the slow-axis polarization is (a) \unit[0.49]{ns}, (b) \unit[2.43]{ns}, and (c) \unit[4.33]{ns}. }
\end{figure}

\begin{figure}
\centering\includegraphics[width=6.5 cm]{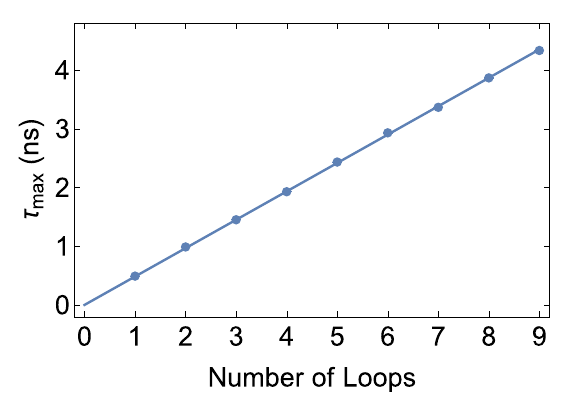}
\caption{\label{fig:delay} The circles are measured differences $\tau_\text{max}$ in mean arrival times between photons polarized along the slow and fast axes of the DGD, plotted as a function of the number of loops. Error bars, corresponding to the standard deviation of the mean, are smaller than the circles. The line is a linear fit to the data that is constrained to pass through the origin.
}
\end{figure}
\subsection{\label{sec:arrival}Arrival times as a function of the number of Zeno stages}

Figure~\ref{fig:1_5_9} shows probabilities of measured photon arrival times for different numbers of loops, and for two polarization settings: polarization along the fast axis of the DGD, and polarization along the slow axis. The relative shift of the center of the photon distribution is clearly seen. The data for one loop verify the weak-measurement regime: The shift of the distribution is much smaller than the width. Figure~\ref{fig:delay} shows measured values for the maximum shift $\tau_\text{max}$, set by the mean arrival time $\bar{t}_\text{ar}(H)$ for photons polarized along the slow axis of the DGD, as a function of the number of loops. 
A linear relationship is observed, and the slope is found to be $\unit[0.483 \pm 0.001]{ns/loop}$, which we use as a calibration for our polarization measurements.

\begin{table}
  \begin{ruledtabular}
    \begin{tabular}{llccccc}
      Distribution & $\tau$ (ns) & $\langle O \rangle$ & $\sigma_\text{PM}$ & $\sigma_\text{SM}$ & $\mathcal{R}$\\ \hline
     (a) (orange circles) & $0.00 \pm 0.79$ & $-1.00$ & 0.41 & 0 & 0\\
    (b) (red diamonds) & $1.07 \pm 0.87$ & $-0.45$ & 0.45 & 0.89 & 2.0\\
     (c) (black asterisks) & $2.07 \pm 0.89$ & $0.07$ & 0.46 & 1.00 & 2.2\\
      (d) (brown triangles) & $3.04 \pm 0.85$ & $0.57$ & 0.44 & 0.82 & 1.9\\
     (e) (blue squares) & $3.87 \pm 0.81$ & $1.00$ & 0.42 & 0 & 0
\end{tabular}
\end{ruledtabular}
\caption{\label{tab:values}Data for the arrival time distributions labeled (a)--(e)  in Fig.~\ref{fig:multiple}. Shown are (i) measured delays $\tau$ relative to the zero defined by the mean arrival time for photons polarized along the fast axis,  with uncertainties given by the standard deviation; (ii) corresponding expectation values $\langle O \rangle$ of linear polarization; (iii) uncertainties $\sigma_\text{PM}$ for the PM obtained from the uncertainties in the arrival times $\tau$, rescaled to the range $[-1,1]$ of the expectation value; (iv) uncertainties $\sigma_\text{SM}$ for the strong measurement obtained from Eq.~\eqref{eq:sigma_SM}; and (v) relative measurement performance $\mathcal{R}=\frac{\sigma_\text{SM}}{\sigma_\text{PM}}$ [see Eq.~\eqref{eq:R}], assuming the same number of detected photons in both the strong and protective measurements.}
 \end{table}

\begin{figure}
\centering\includegraphics[width=8.2 cm]{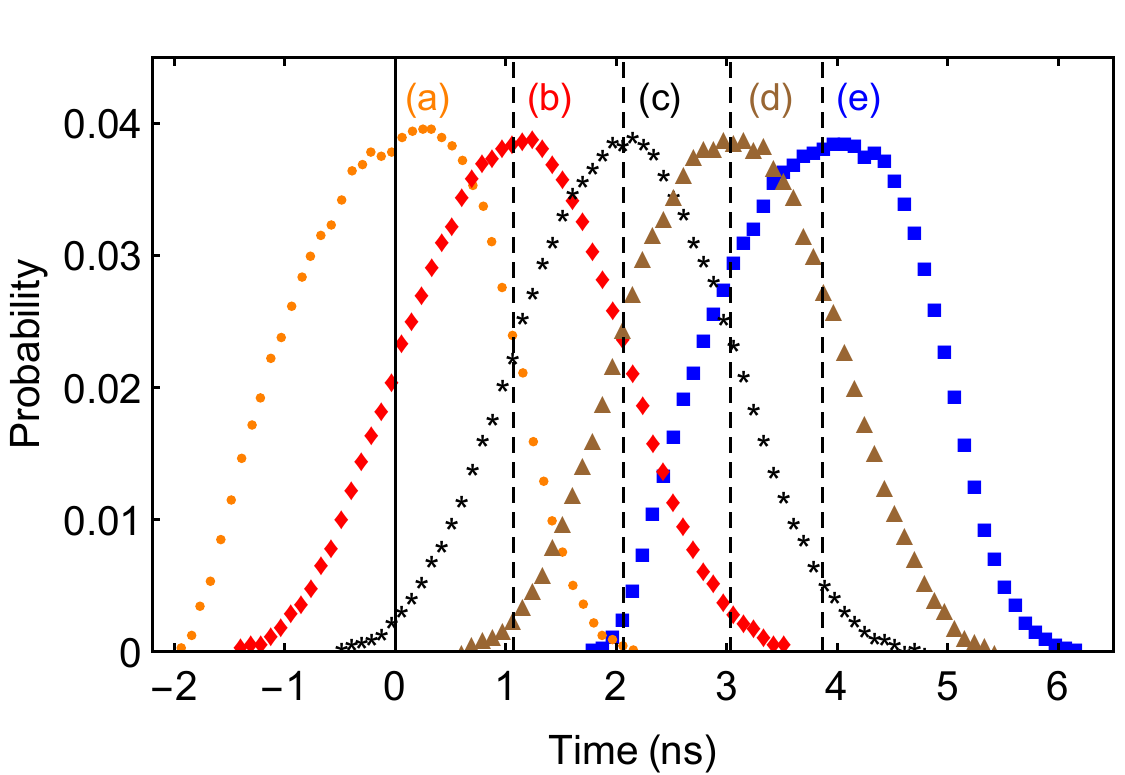}
\caption{\label{fig:multiple} Probabilities of photon arrival times after eight loops. Each curve corresponds to a different polarization. Orange circles (a) correspond to polarization along the fast axis of the DGD, blue squares (e) correspond to polarization along the slow axis, and curves (b)--(d) are polarizations along intermediate angles. The dashed lines correspond to the mean arrival time for each polarization (given in Table~\ref{tab:values}). }
\end{figure}

\subsection{\label{sec:pol}Measuring photon polarization}

Figure~\ref{fig:multiple} shows the measured arrival times of photons after eight loops for five different polarization angles $\theta$. Here, starting from polarization along the fast axis of the DGD, 
we used the PC (Fig.~\ref{fig:app}) to rotate the polarization entering the DGD in steps \footnote{The center {``}paddle{''} in our 3-paddle manual PC behaves approximately as a half-wave plate and therefore rotates the angle $\theta$ in the prepared polarization state. Minor adjustments of the other paddles are used to align the polarization along the fast and slow axes of the DGD.} until we reached the setting in which the photons were polarized along the slow axis. We observe the expected behavior for the PM: The temporal shift consistently increases  as the polarization is rotated from the fast axis to the slow axis, and this shift is therefore indicative of the expectation value of polarization $\langle O \rangle$. (We note that because our experiment did not allow us to independently measure the prepared polarization with any degree of confidence, we  cannot compare $\langle O \rangle$ as obtained from the PM shift to $\langle O \rangle$ as predicted by the input state.) Measured delays are shown in Table~\ref{tab:values}, together with values of $\langle O \rangle$ calculated from these delays by rescaling the delay to the interval $[-1,1]$: $\langle O \rangle = 2\frac{\tau}{\tau_\text{max}}-1$, where $\tau_\text{max} = \unit[(3.864 \pm 0.008)]{ns}$ is the expected delay for polarization along the slow axis of the DGD after eight loops. 


\subsection{\label{sec:uncert}Measurement uncertainties}

It is natural to ask to what extent a Zeno protective measurement may offer advantages (whether practical or fundamental) over a standard strong measurement for determining expectation values. The task of providing a fair comparison is far from straightforward. First, the experimental arrangements are rather different for these two measurement schemes. The prototypical arrangement for a strong measurement may simply consist of a polarizing beamsplitter with a pair of single-photon detectors placed at the two output ports, such that the expectation value can be obtained from the measured photon counts $N_H$ and $N_V$ at those ports via $\langle \op{O} \rangle = (N_H-N_V)/(N_H+N_V)$. A protective-measurement experiment such as ours, or that of Piacentini \emph{et al.\ }\cite{Piacentini:2017:oo}, requires significantly more components than an SM. For example, our experiment uses a $2 \times 2$ switch and a polarization stabilizer, while the experiment of Piacentini \emph{et al.} requires multiple measurement-and-protection stages and a two-dimensional single-photon-detection array. Thus, it is clear that the amount of required experimental resources differs greatly between the SM and PM.

Another resource-based argument that one may consider concerns the required amount of \emph{a priori} information about the photon state for the measurement scheme to work. While for the SM no such information is needed, it may seem that in order to implement the protection, we must have information about the initial photon state since we need to ensure projection onto this initial state at each protection stage. This is arguably indeed the case in the experiment by Piacentini \emph{et al.\ }\cite{Piacentini:2017:oo} since there both preparation and protection are realized through a series of physically separate polarizers, which must be manually dialed to the same setting. Therefore, the prepared (and subsequently protected) state is effectively known to the experimenter. In our experiment, however, the situation is different. As described in Sec.~\ref{sec:app}, not only do we not know what the prepared and protected state is, but the particular experimental arrangement we use also ensures that we \emph{cannot} know this state, even in principle. 
 
An additional important consideration in comparing protective and strong measurements is the issue of intrinsic photon loss. In an ideal SM where optical imperfections can be neglected, there is no such loss: Every photon entering the beamsplitter will also be detected. In the Zeno PM, however, there is a fundamental photon loss even in the absence of optical imperfections because the probability of a photon  surviving the protection stage is less than 1. Previous studies of the performance advantage of protective measurements \cite{Piacentini:2017:oo} focused on the fact that this intrinsic loss is small (see the estimates given in Sec.~\ref{sec:theory}) and that even when taking these intrinsic losses into account, the uncertainty in the expectation value obtained from the PM is smaller than for the SM for the same number of \emph{initial} photons. 

In a practical setting such as ours, however, such intrinsic losses are drowned out by the losses  from the fiber-optic elements in the loop. As stated in Sec.~\ref{sec:theory}, in our experiment  these losses are $\sim\!\unit[10]{dB}$ (90\%) per loop. Thus, if we compared the PM to the SM for the same number of initial photons, the uncertainty of the PM would be far greater simply because the number of detected photons would be so much smaller than in the SM. Alternatively, we may therefore choose to provide a comparison between the PM and SM for the same number of \emph{detected} photons. Such a comparison can be justified by noting that the fiber-optic losses in our experiment are merely practical constraints that may be overcome with an improved experimental implementation. In principle, then, only the intrinsic losses will be left and, as already mentioned, are small enough that they do not significantly influence the photon numbers.

A final difference between the PM and SM concerns the source of the uncertainty in the measured expectation value. In the PM, the pointer shifts deterministically, and the uncertainty is due to the measured width of the pointer wave packet. We can therefore estimate this uncertainty 
 from the  standard deviation $\sigma_\text{PM}$ 
  of the distribution $P_i$ of arrival times (Table~\ref{tab:values}), for a given polarization state and number of loops. While making the pointer wave packet narrower will decrease this uncertainty, there is a trade-off involved because doing so will also make the measurement less weak, since the overlap between the pre- and postmeasurement wave packets will become smaller.

In an ideal PM, the measurement interaction shifts the pointer without changing the pointer wave packet, and the arrival-time distributions in Figs.~\ref{fig:1_5_9} and \ref{fig:multiple} show that this is true to a good approximation also in our experiment. The measured widths $\sigma_\text{PM}$ are similar to the width of the premeasurement wave packet, and are largely independent of the  expectation value $\langle O \rangle$.   

In an SM, the wave packets corresponding to orthogonal polarizations become fully separated, and thus their width and shape are irrelevant. Instead, the  uncertainty in the expectation value 
is now due to the fundamental quantum uncertainty of $\ket{\psi_0}$,
\begin{equation}\label{eq:sigma_SM}
    \sigma_\text{SM}= \sqrt{\langle \op{O}^2\rangle – \langle \op{O} \rangle^2} =\sqrt{1 –\langle \op{O} \rangle^2} =|\sin(2 \theta)|,
\end{equation}
 associated with the standard deviation of the 
linear polarization for the observable $\op{O} = \ketbra{H}{H} - \ketbra{V}{V}$
in the   state $\ket{\psi_0}$. Using Eq.~(\ref{eq:sigma_SM}), we can estimate $\sigma_\text{SM}$  from the measured expectation values (Table~\ref{tab:values}).

Since we use a PS to perform the PM, it is worth considering whether such a stabilizer would improve the performance of an SM. Indeed, such a stabilizer would, in principle, transmit all of the photons through one of the output ports of the polarizing beamsplitter performing the SM, and the uncertainty in the measurement would then be zero. If there are no random phase fluctuations added to the prepared state, reading out the transformation induced by the PS would determine $\langle \op{O} \rangle$ with little or no uncertainty. However, if there is some source of random phase fluctuations that disturb the initial state, as described above in Sec.~\ref{sec:app}, it is not possible to separate this randomness from the state preparation. In this case the SM yields absolutely no information about $\langle \op{O} \rangle$ (the SM is has no uncertainty, but there is no available information about what polarization states are actually being measured by the detectors), so the PM is clearly superior if there is added phase noise.

To compare the PM and SM uncertainties, we shall here take the aforementioned approach of considering the same number of \emph{detected} (rather than initial) photons for both measurement schemes. That is, we  scale both the values $\sigma_\text{PM}$ and $\sigma_\text{SM}$ by $1/\sqrt{N}$ for $N$ recorded photons to get the standard deviation of the mean, $u_\text{PM} = \sigma_\text{PM}/\sqrt{N}$ and $u_\text{SM} = \sigma_\text{SM}/\sqrt{N}$.
Adopting the approach of Ref.~\cite{Piacentini:2017:oo}, we can then assess the measurement performance of the PM relative to the SM as the ratio of the uncertainties for each measurement scheme, i.e.,
\begin{equation}\label{eq:R}
\mathcal{R} = \frac{u_\text{SM}}{u_\text{PM}} = \frac{\sigma_\text{SM}}{\sigma_\text{PM}}=\frac{\sqrt{1 – \langle \op{O} \rangle^2}}{\sigma_\text{PM}}.
\end{equation}
Results are shown in Table~\ref{tab:values}. If $\mathcal{R}  >1$, then the PM provides a lower-uncertainty estimate of the expectation value than the SM, 
given the same number of detected photons for both measurements. 
Table~\ref{tab:values} shows this is the case  
for all input states used in our experiment 
except the ``extreme'' states polarized nearly entirely along the fast or slow axes of the DGD. Of course, for the reasons discussed above, such an observation does not imply the conclusion that our experiment provides a method for measuring expectation values that is superior to an SM. The  experimental resources and the optical losses (which are here effectively ignored by considering only the number of detected photons) for the PM are simply far too great to allow for a fair comparison.



As discussed in Sec.~\ref{sec:theory}, since the uncertainty of the PM is related to the temporal width of the photon wave packet, it may be reduced by decreasing this width. 
However, the  width 
must be kept longer than the amount of DGD per loop to remain in the weak-measurement regime. For a fixed ratio of DGD to pulse width and for a sufficient signal-to-noise ratio (SNR), the performance of the PM increases with the number of loops and is essentially independent of other factors. In our experiment, we are limited to nine loops because of two issues: background and loss. Our high loss rate of $\sim\!\unit[10]{dB}$ per loop (due primarily to losses in the $2 \times 2$ switch, the circulators, and the polarizers) means that not many photons survive large numbers of loops. After eight loops our SNR (defined as the ratio of the number of counts in the peak of the histogram to the number of counts in the background) is  $\sim\!20$, whereas after nine loops it is $\sim\!2$. Background photons come from cw light that makes it through the $1 \times 1$ switches (due to finite contrast of those switches) and from the cw light used to stabilize the polarization. It may be possible to find another way to stabilize the polarization without using cw light, and this would eliminate a source of background, as well as eliminate the need for the circulators. It may also be possible to find lower-loss polarizers and switches. This would allow us to increase the number of loops and  improve the performance of our measurements.

\section{\label{sec:conclusions}Concluding discussion}

We have demonstrated  an experimental implementation of  a protective measurement based on the quantum Zeno effect. Our approach is different in two ways: We use a temporal pointer in the form of polarization-dependent photon arrival times, and we implement a loop configuration in which we let the photon repeatedly cycle through a Zeno stage for up to nine stages. This approach offers enhanced flexibility because the number of stages can be adjusted easily without having to physically modify the arrangement of the optical elements.

We have experimentally confirmed the theoretically expected behavior for a PM, namely, that the pointer wave packet (represented by  the distribution of recorded photon arrival times) shifts proportionally to the number of loops (Figs.~\ref{fig:1_5_9} and \ref{fig:delay}) and that for a given number of loops this shift changes in response to changes in the polarization state (Fig.~\ref{fig:multiple}). 

A noteworthy feature of our experiment is the fact that no knowledge of the initial polarization state of the photon is required to realize the state protection, which is implemented using a  polarization stabilizer. Moreover, not only do we not need to know the state in order to protect it, but indeed it is fundamentally impossible in our experiment for the experimenter to know what the state is. 
Thus, while on the surface it may appear as though one would always need to know what the state is in order to protect it, our experiment shows that this is not so. Because no state information is available, the expectation values measured in the Zeno PM indeed yield fresh information. 

\begin{acknowledgments}
We thank J.~Ewing for help with construction of the experimental apparatus. This work was funded by the National Science Foundation (Grant No.\ PHY-2109964/2109962). M.-W.\,C. acknowledges support from the Gordon and Betty Moore Foundation.
\end{acknowledgments}


\begin{thebibliography}{24}%
\makeatletter
\providecommand \@ifxundefined [1]{%
 \@ifx{#1\undefined}
}%
\providecommand \@ifnum [1]{%
 \ifnum #1\expandafter \@firstoftwo
 \else \expandafter \@secondoftwo
 \fi
}%
\providecommand \@ifx [1]{%
 \ifx #1\expandafter \@firstoftwo
 \else \expandafter \@secondoftwo
 \fi
}%
\providecommand \natexlab [1]{#1}%
\providecommand \enquote  [1]{``#1''}%
\providecommand \bibnamefont  [1]{#1}%
\providecommand \bibfnamefont [1]{#1}%
\providecommand \citenamefont [1]{#1}%
\providecommand \href@noop [0]{\@secondoftwo}%
\providecommand \href [0]{\begingroup \@sanitize@url \@href}%
\providecommand \@href[1]{\@@startlink{#1}\@@href}%
\providecommand \@@href[1]{\endgroup#1\@@endlink}%
\providecommand \@sanitize@url [0]{\catcode `\\12\catcode `\$12\catcode
  `\&12\catcode `\#12\catcode `\^12\catcode `\_12\catcode `\%12\relax}%
\providecommand \@@startlink[1]{}%
\providecommand \@@endlink[0]{}%
\providecommand \url  [0]{\begingroup\@sanitize@url \@url }%
\providecommand \@url [1]{\endgroup\@href {#1}{\urlprefix }}%
\providecommand \urlprefix  [0]{URL }%
\providecommand \Eprint [0]{\href }%
\providecommand \doibase [0]{https://doi.org/}%
\providecommand \selectlanguage [0]{\@gobble}%
\providecommand \bibinfo  [0]{\@secondoftwo}%
\providecommand \bibfield  [0]{\@secondoftwo}%
\providecommand \translation [1]{[#1]}%
\providecommand \BibitemOpen [0]{}%
\providecommand \bibitemStop [0]{}%
\providecommand \bibitemNoStop [0]{.\EOS\space}%
\providecommand \EOS [0]{\spacefactor3000\relax}%
\providecommand \BibitemShut  [1]{\csname bibitem#1\endcsname}%
\let\auto@bib@innerbib\@empty
\bibitem [{\citenamefont {Aharonov}\ \emph {et~al.}(1988)\citenamefont
  {Aharonov}, \citenamefont {Albert},\ and\ \citenamefont
  {Vaidman}}]{Aharonov:1988:mz}%
  \BibitemOpen
  \bibfield  {author} {\bibinfo {author} {\bibfnamefont {Y.}~\bibnamefont
  {Aharonov}}, \bibinfo {author} {\bibfnamefont {D.~Z.}\ \bibnamefont
  {Albert}},\ and\ \bibinfo {author} {\bibfnamefont {L.}~\bibnamefont
  {Vaidman}},\ }\bibfield  {title} {\bibinfo {title} {How the result of a
  measurement of a component of the spin of a spin-1/2 particle can turn out to
  be 100},\ }\href {https://doi.org/10.1103/PhysRevLett.60.1351} {\bibfield
  {journal} {\bibinfo  {journal} {Phys. Rev. Lett.}\ }\textbf {\bibinfo
  {volume} {60}},\ \bibinfo {pages} {1351} (\bibinfo {year}
  {1988})}\BibitemShut {NoStop}%
\bibitem [{\citenamefont {Tamir}\ and\ \citenamefont
  {Cohen}(2013)}]{Tamir:2013:za}%
  \BibitemOpen
  \bibfield  {author} {\bibinfo {author} {\bibfnamefont {B.}~\bibnamefont
  {Tamir}}\ and\ \bibinfo {author} {\bibfnamefont {E.}~\bibnamefont {Cohen}},\
  }\bibfield  {title} {\bibinfo {title} {Introduction to weak measurements and
  weak values},\ }\href {https://doi.org/10.12743/quanta.v2i1.14} {\bibfield
  {journal} {\bibinfo  {journal} {Quanta}\ }\textbf {\bibinfo {volume} {2}},\
  \bibinfo {pages} {7} (\bibinfo {year} {2013})}\BibitemShut {NoStop}%
\bibitem [{\citenamefont {Dressel}\ \emph {et~al.}(2014)\citenamefont
  {Dressel}, \citenamefont {Malik}, \citenamefont {Miatto}, \citenamefont
  {Jordan},\ and\ \citenamefont {Boyd}}]{Dressel:2014:uu}%
  \BibitemOpen
  \bibfield  {author} {\bibinfo {author} {\bibfnamefont {J.}~\bibnamefont
  {Dressel}}, \bibinfo {author} {\bibfnamefont {M.}~\bibnamefont {Malik}},
  \bibinfo {author} {\bibfnamefont {F.~M.}\ \bibnamefont {Miatto}}, \bibinfo
  {author} {\bibfnamefont {A.~N.}\ \bibnamefont {Jordan}},\ and\ \bibinfo
  {author} {\bibfnamefont {R.~W.}\ \bibnamefont {Boyd}},\ }\bibfield  {title}
  {\bibinfo {title} {Colloquium: Understanding quantum weak values: Basics and
  applications},\ }\href {https://doi.org/10.1103/RevModPhys.86.307} {\bibfield
   {journal} {\bibinfo  {journal} {Rev. Mod. Phys.}\ }\textbf {\bibinfo
  {volume} {86}},\ \bibinfo {pages} {307} (\bibinfo {year} {2014})}\BibitemShut
  {NoStop}%
\bibitem [{\citenamefont {Arvidsson-Shukur}\ \emph {et~al.}(2020)\citenamefont
  {Arvidsson-Shukur}, \citenamefont {Yunger~Halpern}, \citenamefont {Lepage},
  \citenamefont {Lasek}, \citenamefont {Barnes},\ and\ \citenamefont
  {Lloyd}}]{ArvidssonShukur:2020:az}%
  \BibitemOpen
  \bibfield  {author} {\bibinfo {author} {\bibfnamefont {D.~R.~M.}\
  \bibnamefont {Arvidsson-Shukur}}, \bibinfo {author} {\bibfnamefont
  {N.}~\bibnamefont {Yunger~Halpern}}, \bibinfo {author} {\bibfnamefont
  {H.~V.}\ \bibnamefont {Lepage}}, \bibinfo {author} {\bibfnamefont {A.~A.}\
  \bibnamefont {Lasek}}, \bibinfo {author} {\bibfnamefont {C.~H.~W.}\
  \bibnamefont {Barnes}},\ and\ \bibinfo {author} {\bibfnamefont
  {S.}~\bibnamefont {Lloyd}},\ }\bibfield  {title} {\bibinfo {title} {Quantum
  advantage in postselected metrology},\ }\href
  {https://doi.org/10.1038/s41467-020-17559-w} {\bibfield  {journal} {\bibinfo
  {journal} {Nature Comm.}\ }\textbf {\bibinfo {volume} {11}},\ \bibinfo
  {pages} {3775} (\bibinfo {year} {2020})}\BibitemShut {NoStop}%
\bibitem [{\citenamefont {Aharonov}\ and\ \citenamefont
  {Vaidman}(1993)}]{Aharonov:1993:qa}%
  \BibitemOpen
  \bibfield  {author} {\bibinfo {author} {\bibfnamefont {Y.}~\bibnamefont
  {Aharonov}}\ and\ \bibinfo {author} {\bibfnamefont {L.}~\bibnamefont
  {Vaidman}},\ }\bibfield  {title} {\bibinfo {title} {Measurement of the
  {S}chr{\"o}dinger wave of a single particle},\ }\href
  {https://doi.org/10.1016/0375-9601(93)90724-E} {\bibfield  {journal}
  {\bibinfo  {journal} {Phys. Lett. A}\ }\textbf {\bibinfo {volume} {178}},\
  \bibinfo {pages} {38} (\bibinfo {year} {1993})}\BibitemShut {NoStop}%
\bibitem [{\citenamefont {Aharonov}\ \emph {et~al.}(1993)\citenamefont
  {Aharonov}, \citenamefont {Anandan},\ and\ \citenamefont
  {Vaidman}}]{Aharonov:1993:jm}%
  \BibitemOpen
  \bibfield  {author} {\bibinfo {author} {\bibfnamefont {Y.}~\bibnamefont
  {Aharonov}}, \bibinfo {author} {\bibfnamefont {J.}~\bibnamefont {Anandan}},\
  and\ \bibinfo {author} {\bibfnamefont {L.}~\bibnamefont {Vaidman}},\
  }\bibfield  {title} {\bibinfo {title} {Meaning of the wave function},\ }\href
  {https://doi.org/10.1103/PhysRevA.47.4616} {\bibfield  {journal} {\bibinfo
  {journal} {Phys. Rev. A}\ }\textbf {\bibinfo {volume} {47}},\ \bibinfo
  {pages} {4616} (\bibinfo {year} {1993})}\BibitemShut {NoStop}%
\bibitem [{\citenamefont {Anandan}(1993)}]{Anandan:1993:uu}%
  \BibitemOpen
  \bibfield  {author} {\bibinfo {author} {\bibfnamefont {J.}~\bibnamefont
  {Anandan}},\ }\bibfield  {title} {\bibinfo {title} {Protective measurement
  and quantum reality},\ }\href {https://doi.org/10.1007/BF00662803} {\bibfield
   {journal} {\bibinfo  {journal} {Found. Phys. Lett.}\ }\textbf {\bibinfo
  {volume} {6}},\ \bibinfo {pages} {503} (\bibinfo {year} {1993})}\BibitemShut
  {NoStop}%
\bibitem [{\citenamefont {{Hari Dass}}\ and\ \citenamefont
  {Qureshi}(1999)}]{Dass:1999:az}%
  \BibitemOpen
  \bibfield  {author} {\bibinfo {author} {\bibfnamefont {N.~D.}\ \bibnamefont
  {{Hari Dass}}}\ and\ \bibinfo {author} {\bibfnamefont {T.}~\bibnamefont
  {Qureshi}},\ }\bibfield  {title} {\bibinfo {title} {Critique of protective
  measurements},\ }\href {https://doi.org/10.1103/PhysRevA.59.2590} {\bibfield
  {journal} {\bibinfo  {journal} {Phys. Rev. A}\ }\textbf {\bibinfo {volume}
  {59}},\ \bibinfo {pages} {2590} (\bibinfo {year} {1999})}\BibitemShut
  {NoStop}%
\bibitem [{\citenamefont {Vaidman}(2009)}]{Vaidman:2009:po}%
  \BibitemOpen
  \bibfield  {author} {\bibinfo {author} {\bibfnamefont {L.}~\bibnamefont
  {Vaidman}},\ }\bibfield  {title} {\bibinfo {title} {Protective
  measurements},\ }in\ \href@noop {} {\emph {\bibinfo {booktitle} {Compendium
  of Quantum Physics: Concepts, Experiments, History and Philosophy}}},\
  \bibinfo {editor} {edited by\ \bibinfo {editor} {\bibfnamefont
  {D.}~\bibnamefont {Greenberger}}, \bibinfo {editor} {\bibfnamefont
  {K.}~\bibnamefont {Hentschel}},\ and\ \bibinfo {editor} {\bibfnamefont
  {F.}~\bibnamefont {Weinert}}}\ (\bibinfo  {publisher} {Springer},\ \bibinfo
  {address} {Berlin},\ \bibinfo {year} {2009}), pp.\ \bibinfo
  {pages} {505--508}\BibitemShut {NoStop}%
\bibitem [{\citenamefont {Gao}(2014)}]{Gao:2014:cu}%
  \BibitemOpen
  {\emph {\bibinfo {title} {Protective Measurement and Quantum Reality:
  Towards a New Understanding of Quantum Mechanics}}}, edited by S.~Gao (\bibinfo  {publisher}
  {Cambridge University Press},\ \bibinfo {address} {Cambridge},\ \bibinfo
  {year} {2014})\BibitemShut {NoStop}%
\bibitem [{\citenamefont {Genovese}(2017)}]{Genovese:2017:zz}%
  \BibitemOpen
  \bibfield  {author} {\bibinfo {author} {\bibfnamefont {M.}~\bibnamefont
  {Genovese}},\ }\bibfield  {title} {\bibinfo {title} {A few reflections on
  protective measurements and more},\ }\href
  {https://doi.org/10.1088/1742-6596/880/1/012012} {\bibfield  {journal}
  {\bibinfo  {journal} {J. Phys.: Conf. Ser.}\ }\textbf {\bibinfo {volume}
  {880}},\ \bibinfo {pages} {012012} (\bibinfo {year} {2017})}\BibitemShut
  {NoStop}%
\bibitem [{\citenamefont {Piacentini}\ \emph {et~al.}(2017)\citenamefont
  {Piacentini}, \citenamefont {Avella}, \citenamefont {Rebufello},
  \citenamefont {Lussana}, \citenamefont {Villa}, \citenamefont {Tosi},
  \citenamefont {Gramegna}, \citenamefont {Brida}, \citenamefont {Cohen},
  \citenamefont {Vaidman}, \citenamefont {Degiovanni},\ and\ \citenamefont
  {Genovese}}]{Piacentini:2017:oo}%
  \BibitemOpen
  \bibfield  {author} {\bibinfo {author} {\bibfnamefont {F.}~\bibnamefont
  {Piacentini}}, \bibinfo {author} {\bibfnamefont {A.}~\bibnamefont {Avella}},
  \bibinfo {author} {\bibfnamefont {E.}~\bibnamefont {Rebufello}}, \bibinfo
  {author} {\bibfnamefont {R.}~\bibnamefont {Lussana}}, \bibinfo {author}
  {\bibfnamefont {F.}~\bibnamefont {Villa}}, \bibinfo {author} {\bibfnamefont
  {A.}~\bibnamefont {Tosi}}, \bibinfo {author} {\bibfnamefont {M.}~\bibnamefont
  {Gramegna}}, \bibinfo {author} {\bibfnamefont {G.}~\bibnamefont {Brida}},
  \bibinfo {author} {\bibfnamefont {E.}~\bibnamefont {Cohen}}, \bibinfo
  {author} {\bibfnamefont {L.}~\bibnamefont {Vaidman}}, \bibinfo {author}
  {\bibfnamefont {I.~P.}\ \bibnamefont {Degiovanni}},\ and\ \bibinfo {author}
  {\bibfnamefont {M.}~\bibnamefont {Genovese}},\ }\bibfield  {title} {\bibinfo
  {title} {Determining the quantum expectation value by measuring a single
  photon},\ }\href {https://doi.org/10.1038/nphys4223} {\bibfield  {journal}
  {\bibinfo  {journal} {Nature Phys.}\ }\textbf {\bibinfo {volume} {13}},\
  \bibinfo {pages} {1191} (\bibinfo {year} {2017})}\BibitemShut {NoStop}%
\bibitem [{\citenamefont {Rebufello}\ \emph {et~al.}(2021)\citenamefont
  {Rebufello}, \citenamefont {Piacentini}, \citenamefont {Avella},
  \citenamefont {Lussana}, \citenamefont {Villa}, \citenamefont {Tosi},
  \citenamefont {Gramegna}, \citenamefont {Brida}, \citenamefont {Cohen},
  \citenamefont {Vaidman}, \citenamefont {Degiovanni},\ and\ \citenamefont
  {Genovese}}]{rebufello_2021}%
  \BibitemOpen
  \bibfield  {author} {\bibinfo {author} {\bibfnamefont {E.}~\bibnamefont
  {Rebufello}}, \bibinfo {author} {\bibfnamefont {F.}~\bibnamefont
  {Piacentini}}, \bibinfo {author} {\bibfnamefont {A.}~\bibnamefont {Avella}},
  \bibinfo {author} {\bibfnamefont {R.}~\bibnamefont {Lussana}}, \bibinfo
  {author} {\bibfnamefont {F.}~\bibnamefont {Villa}}, \bibinfo {author}
  {\bibfnamefont {A.}~\bibnamefont {Tosi}}, \bibinfo {author} {\bibfnamefont
  {M.}~\bibnamefont {Gramegna}}, \bibinfo {author} {\bibfnamefont
  {G.}~\bibnamefont {Brida}}, \bibinfo {author} {\bibfnamefont
  {E.}~\bibnamefont {Cohen}}, \bibinfo {author} {\bibfnamefont
  {L.}~\bibnamefont {Vaidman}}, \bibinfo {author} {\bibfnamefont {I.~P.}\
  \bibnamefont {Degiovanni}},\ and\ \bibinfo {author} {\bibfnamefont
  {M.}~\bibnamefont {Genovese}},\ }\bibfield  {title} {\bibinfo {title}
  {Protective measurement\textemdash{A new quantum measurement paradigm}:
  {D}etailed description of the first realization},\ }\href
  {https://doi.org/10.3390/app11094260} {\bibfield  {journal} {\bibinfo
  {journal} {Appl. Sci.}\ }\textbf {\bibinfo {volume} {11}},\ \bibinfo {pages}
  {4260} (\bibinfo {year} {2021})}\BibitemShut {NoStop}%
\bibitem [{\citenamefont {Zhang}\ and\ \citenamefont
  {Gong}(2020)}]{Zhang:2020:aa}%
  \BibitemOpen
  \bibfield  {author} {\bibinfo {author} {\bibfnamefont {D.-J.}\ \bibnamefont
  {Zhang}}\ and\ \bibinfo {author} {\bibfnamefont {J.}~\bibnamefont {Gong}},\
  }\bibfield  {title} {\bibinfo {title} {Dissipative adiabatic measurements:
  Beating the quantum {C}ram{\'e}r--{R}ao bound},\ }\href
  {https://doi.org/10.1103/PhysRevResearch.2.023418} {\bibfield  {journal}
  {\bibinfo  {journal} {Phys. Rev. Research}\ }\textbf {\bibinfo {volume}
  {2}},\ \bibinfo {pages} {023418} (\bibinfo {year} {2020})}\BibitemShut
  {NoStop}%
\bibitem [{\citenamefont {Misra}\ and\ \citenamefont
  {Sudarshan}(1977)}]{Misra:1977:aa}%
  \BibitemOpen
  \bibfield  {author} {\bibinfo {author} {\bibfnamefont {B.}~\bibnamefont
  {Misra}}\ and\ \bibinfo {author} {\bibfnamefont {E.~C.~G.}\ \bibnamefont
  {Sudarshan}},\ }\bibfield  {title} {\bibinfo {title} {The {Z}eno's paradox in
  quantum theory},\ }\href {https://doi.org/10.1063/1.523304} {\bibfield
  {journal} {\bibinfo  {journal} {J. Math. Phys.}\ }\textbf {\bibinfo {volume}
  {18}},\ \bibinfo {pages} {756} (\bibinfo {year} {1977})}\BibitemShut
  {NoStop}%
\bibitem [{\citenamefont {Itano}\ \emph {et~al.}(1990)\citenamefont {Itano},
  \citenamefont {Heinzen}, \citenamefont {Bollinger},\ and\ \citenamefont
  {Wineland}}]{Itano:1990:lm}%
  \BibitemOpen
  \bibfield  {author} {\bibinfo {author} {\bibfnamefont {W.~M.}\ \bibnamefont
  {Itano}}, \bibinfo {author} {\bibfnamefont {D.~J.}\ \bibnamefont {Heinzen}},
  \bibinfo {author} {\bibfnamefont {J.~J.}\ \bibnamefont {Bollinger}},\ and\
  \bibinfo {author} {\bibfnamefont {D.~J.}\ \bibnamefont {Wineland}},\
  }\bibfield  {title} {\bibinfo {title} {Quantum {Z}eno effect},\ }\href
  {https://doi.org/10.1103/PhysRevA.41.2295} {\bibfield  {journal} {\bibinfo
  {journal} {Phys. Rev. A}\ }\textbf {\bibinfo {volume} {41}},\ \bibinfo
  {pages} {2295} (\bibinfo {year} {1990})}\BibitemShut {NoStop}%
\bibitem [{\citenamefont {Home}\ and\ \citenamefont
  {Whitaker}(1997)}]{Home:1997:za}%
  \BibitemOpen
  \bibfield  {author} {\bibinfo {author} {\bibfnamefont {D.}~\bibnamefont
  {Home}}\ and\ \bibinfo {author} {\bibfnamefont {M.}~\bibnamefont
  {Whitaker}},\ }\bibfield  {title} {\bibinfo {title} {A conceptual analysis of
  quantum {Z}eno; paradox, measurement, and experiment},\ }\href
  {https://doi.org/https://doi.org/10.1006/aphy.1997.5699} {\bibfield
  {journal} {\bibinfo  {journal} {Ann. Phys. (NY)}\ }\textbf {\bibinfo {volume}
  {258}},\ \bibinfo {pages} {237} (\bibinfo {year} {1997})}\BibitemShut
  {NoStop}%
\bibitem [{\citenamefont {Virz\`{\i}}\ \emph {et~al.}(2022)\citenamefont
  {Virz\`{\i}}, \citenamefont {Avella}, \citenamefont {Piacentini},
  \citenamefont {Gramegna}, \citenamefont {Opatrn\'y}, \citenamefont {Kofman},
  \citenamefont {Kurizki}, \citenamefont {Gherardini}, \citenamefont {Caruso},
  \citenamefont {Degiovanni},\ and\ \citenamefont {Genovese}}]{Virzi:2022:aa}%
  \BibitemOpen
  \bibfield  {author} {\bibinfo {author} {\bibfnamefont {S.}~\bibnamefont
  {Virz\`{\i}}}, \bibinfo {author} {\bibfnamefont {A.}~\bibnamefont {Avella}},
  \bibinfo {author} {\bibfnamefont {F.}~\bibnamefont {Piacentini}}, \bibinfo
  {author} {\bibfnamefont {M.}~\bibnamefont {Gramegna}}, \bibinfo {author}
  {\bibfnamefont {T.}\ \bibnamefont {Opatrn\'y}}, \bibinfo {author}
  {\bibfnamefont {A.~G.}\ \bibnamefont {Kofman}}, \bibinfo {author}
  {\bibfnamefont {G.}~\bibnamefont {Kurizki}}, \bibinfo {author} {\bibfnamefont
  {S.}~\bibnamefont {Gherardini}}, \bibinfo {author} {\bibfnamefont
  {F.}~\bibnamefont {Caruso}}, \bibinfo {author} {\bibfnamefont {I.~P.}\
  \bibnamefont {Degiovanni}},\ and\ \bibinfo {author} {\bibfnamefont
  {M.}~\bibnamefont {Genovese}},\ }\bibfield  {title} {\bibinfo {title}
  {Quantum {Z}eno and anti-{Z}eno probes of noise correlations in photon
  polarization},\ }\href {https://doi.org/10.1103/PhysRevLett.129.030401}
  {\bibfield  {journal} {\bibinfo  {journal} {Phys. Rev. Lett.}\ }\textbf
  {\bibinfo {volume} {129}},\ \bibinfo {pages} {030401} (\bibinfo {year}
  {2022})}\BibitemShut {NoStop}%
\bibitem [{\citenamefont {Schlosshauer}(2018)}]{Schlosshauer:2018:xx}%
  \BibitemOpen
  \bibfield  {author} {\bibinfo {author} {\bibfnamefont {M.}~\bibnamefont
  {Schlosshauer}},\ }\bibfield  {title} {\bibinfo {title} {Scheme for the
  protective measurement of a single photon using a tunable quantum {Z}eno
  effect},\ }\href {https://doi.org/10.1103/PhysRevA.97.042104} {\bibfield
  {journal} {\bibinfo  {journal} {Phys. Rev. A}\ }\textbf {\bibinfo {volume}
  {97}},\ \bibinfo {pages} {042104} (\bibinfo {year} {2018})}\BibitemShut
  {NoStop}%
\bibitem [{\citenamefont {Brunner}\ \emph {et~al.}(2003)\citenamefont
  {Brunner}, \citenamefont {Ac\'{\i}n}, \citenamefont {Collins}, \citenamefont
  {Gisin},\ and\ \citenamefont {Scarani}}]{Brunner:2003:az}%
  \BibitemOpen
  \bibfield  {author} {\bibinfo {author} {\bibfnamefont {N.}~\bibnamefont
  {Brunner}}, \bibinfo {author} {\bibfnamefont {A.}~\bibnamefont {Ac\'{\i}n}},
  \bibinfo {author} {\bibfnamefont {D.}~\bibnamefont {Collins}}, \bibinfo
  {author} {\bibfnamefont {N.}~\bibnamefont {Gisin}},\ and\ \bibinfo {author}
  {\bibfnamefont {V.}~\bibnamefont {Scarani}},\ }\bibfield  {title} {\bibinfo
  {title} {Optical telecom networks as weak quantum measurements with
  postselection},\ }\href {https://doi.org/10.1103/PhysRevLett.91.180402}
  {\bibfield  {journal} {\bibinfo  {journal} {Phys. Rev. Lett.}\ }\textbf
  {\bibinfo {volume} {91}},\ \bibinfo {pages} {180402} (\bibinfo {year}
  {2003})}\BibitemShut {NoStop}%
\bibitem [{\citenamefont {Brunner}\ \emph {et~al.}(2004)\citenamefont
  {Brunner}, \citenamefont {Scarani}, \citenamefont {Wegm\"uller},
  \citenamefont {Legr\'e},\ and\ \citenamefont {Gisin}}]{Brunner:2004:aa}%
  \BibitemOpen
  \bibfield  {author} {\bibinfo {author} {\bibfnamefont {N.}~\bibnamefont
  {Brunner}}, \bibinfo {author} {\bibfnamefont {V.}~\bibnamefont {Scarani}},
  \bibinfo {author} {\bibfnamefont {M.}~\bibnamefont {Wegm\"uller}}, \bibinfo
  {author} {\bibfnamefont {M.}~\bibnamefont {Legr\'e}},\ and\ \bibinfo {author}
  {\bibfnamefont {N.}~\bibnamefont {Gisin}},\ }\bibfield  {title} {\bibinfo
  {title} {Direct measurement of superluminal group velocity and signal
  velocity in an optical fiber},\ }\href
  {https://doi.org/10.1103/PhysRevLett.93.203902} {\bibfield  {journal}
  {\bibinfo  {journal} {Phys. Rev. Lett.}\ }\textbf {\bibinfo {volume} {93}},\
  \bibinfo {pages} {203902} (\bibinfo {year} {2004})}\BibitemShut {NoStop}%
\bibitem [{\citenamefont {Wang}\ \emph {et~al.}(2006)\citenamefont {Wang},
  \citenamefont {Sun}, \citenamefont {Zhang}, \citenamefont {Jian-Li},
  \citenamefont {Huang},\ and\ \citenamefont {Guo}}]{Wang:2006:un}%
  \BibitemOpen
  \bibfield  {author} {\bibinfo {author} {\bibfnamefont {Q.}~\bibnamefont
  {Wang}}, \bibinfo {author} {\bibfnamefont {F.-W.}\ \bibnamefont {Sun}},
  \bibinfo {author} {\bibfnamefont {Y.-S.}\ \bibnamefont {Zhang}}, \bibinfo
  {author} {\bibnamefont {Jian-Li}}, \bibinfo {author} {\bibfnamefont {Y.-F.}\
  \bibnamefont {Huang}},\ and\ \bibinfo {author} {\bibfnamefont {G.-C.}\
  \bibnamefont {Guo}},\ }\bibfield  {title} {\bibinfo {title} {Experimental
  demonstration of a method to realize weak measurement of the arrival time of
  a single photon},\ }\href {https://doi.org/10.1103/PhysRevA.73.023814}
  {\bibfield  {journal} {\bibinfo  {journal} {Phys. Rev. A}\ }\textbf {\bibinfo
  {volume} {73}},\ \bibinfo {pages} {023814} (\bibinfo {year}
  {2006})}\BibitemShut {NoStop}%
\bibitem [{\citenamefont {Piacentini}\ \emph {et~al.}(2018)\citenamefont
  {Piacentini}, \citenamefont {Avella}, \citenamefont {Gramegna}, \citenamefont
  {Lussana}, \citenamefont {Villa}, \citenamefont {Tosi}, \citenamefont
  {Brida}, \citenamefont {Degiovanni},\ and\ \citenamefont
  {Genovese}}]{Piacentini:2018:za}%
  \BibitemOpen
  \bibfield  {author} {\bibinfo {author} {\bibfnamefont {F.}~\bibnamefont
  {Piacentini}}, \bibinfo {author} {\bibfnamefont {A.}~\bibnamefont {Avella}},
  \bibinfo {author} {\bibfnamefont {M.}~\bibnamefont {Gramegna}}, \bibinfo
  {author} {\bibfnamefont {R.}~\bibnamefont {Lussana}}, \bibinfo {author}
  {\bibfnamefont {F.}~\bibnamefont {Villa}}, \bibinfo {author} {\bibfnamefont
  {A.}~\bibnamefont {Tosi}}, \bibinfo {author} {\bibfnamefont {G.}~\bibnamefont
  {Brida}}, \bibinfo {author} {\bibfnamefont {I.~P.}\ \bibnamefont
  {Degiovanni}},\ and\ \bibinfo {author} {\bibfnamefont {M.}~\bibnamefont
  {Genovese}},\ }\bibfield  {title} {\bibinfo {title} {Investigating the
  effects of the interaction intensity in a weak measurement},\ }\href
  {https://doi.org/10.1038/s41598-018-25156-7} {\bibfield  {journal} {\bibinfo
  {journal} {Sci. Rep.}\ }\textbf {\bibinfo {volume} {8}},\ \bibinfo {pages}
  {6959} (\bibinfo {year} {2018})}\BibitemShut {NoStop}%
\bibitem [{Note1()}]{Note1}%
  \BibitemOpen
  \bibinfo {note} {The center {``}paddle{''} in our three-paddle manual PC behaves
  approximately as a half-wave plate and therefore rotates the angle $\theta $
  in the prepared polarization state. Minor adjustments of the other paddles
  are used to align the polarization along the fast and slow axes of the
  DGD.}\BibitemShut {Stop}%
\end{thebibliography}
%

\end{document}